\documentclass[a4paper,11pt]{article}

\pdfoutput=1

\usepackage{amsmath,amssymb,mathtools}
\usepackage{color}
\usepackage{graphicx}
\usepackage{subfigure}
\usepackage{cite}
\usepackage[colorlinks=true,linkcolor=blue, citecolor=red, urlcolor=blue, bookmarks]{hyperref}
\usepackage{multirow,makecell} %% multirow and multicolumn
\usepackage[figuresright]{rotating} %% rotating tables
\usepackage{textcomp}
\usepackage{wasysym}
\usepackage{ulem}

\usepackage[utf8]{inputenc}
\usepackage[T1]{fontenc}

\usepackage[text={17.1cm,24.6cm},centering]{geometry} %% thanks to Chao Wu

\numberwithin{equation}{section}

\def \be {\begin{equation}}
\def \ee {\end{equation}}
\def \ba {\begin{array}}
\def \ea {\end{array}}
\def \bea{\begin{eqnarray}}
\def \eea{\end{eqnarray}}
\def \nn {\nonumber}

\def \a {\alpha}
\def \b {\beta}
\def \g {\gamma}

\def \d {\delta}
\def \D {\Delta}

\def \k {\kappa}
\def \l {\lambda}

\def \s {\sigma}

\def \r {\rho}

\def \th {\theta}

\def \t {\tau}
\def \z {\zeta}

\def \cH {\mathcal H}

\def \cL {\mathcal L}
\def \cM {\mathcal M}

\def \cO {\mathcal O}
\def \cP {\mathcal P}
\def \cQ {\mathcal Q}

\def \cS {\mathcal S}

\def \f {\frac}

\def \lt {\left}
\def \rt {\right}

\def \ua {\uparrow}

\def \sr {\sqrt}
\def \td {\tilde}

\def \pp {\propto}

\def \lag {\langle}
\def \rag {\rangle}

\def \ii {\mathrm{i}}

\def \tr {\textrm{tr}}

\def \and {{~\textrm{and}~}}

\def \ibl {\textit{\textbf{l}}}

\begin{document}

\title{
\textbf{Bootstrapping entanglement in quantum spin systems}
}
\author{
Jiaju Zhang$^{1}$, %\footnote{jiajuzhang@tju.edu.cn},
Arash Jafarizadeh$^{2}$ %, %\footnote{arash.jafarizadeh@nottingham.ac.uk},
and
M. A. Rajabpour$^{3}$ %\footnote{mohammadali.rajabpour@gmail.com}
}
\date{}
\maketitle
\vspace{-10mm}
\begin{center}
{\it
$^{1}$Center for Joint Quantum Studies and Department of Physics,\\School of Science, Tianjin University, 135 Yaguan Road, Tianjin 300350, China\\\vspace{1mm}
$^{2}$School of Physics and Astronomy, University of Nottingham,\\University Park, Nottingham NG7 2RD, United Kingdom\\\vspace{1mm}
$^{3}$Instituto de Fisica, Universidade Federal Fluminense,\\
      Av. Gal. Milton Tavares de Souza s/n, Gragoat\'a, 24210-346, Niter\'oi, RJ, Brazil
}
\vspace{10mm}
\end{center}

\begin{abstract}

  %This paper aims to redefine how we understand and calculate the properties of quantum many-body systems, specifically focusing on entanglement in quantum spin systems. Traditional approaches necessitate the diagonalization of the system's Hamiltonian to ascertain properties such as eigenvalues, correlation functions, and quantum entanglement.
  %In contrast, we employ the bootstrap method, a technique that leverages consistency relations rather than direct diagonalization, to estimate these properties.
  In this paper, we employ the bootstrap method, a technique that relies on consistency relations instead of direct diagonalization, to determine the expectation values in quantum many-body systems. We then use these values to assess the entanglement content of the system. Our work extends the bootstrap approach to quantum many-body systems, rather than single-body or few-body systems, concentrating on the well-known Lipkin-Meshkov-Glick (LMG) model with both transverse and longitudinal external magnetic fields. In the bootstrap method we solve the LMG model with up to 16 sites. Unlike previous studies that have focused mainly on ground-state properties, our methodology allows for the calculation of a broad range of properties, including energy spectrum, angular momentum, concurrence, tangle, residual tangle, and quantum Fisher information (QFI), for all eigenstates or a particular sector of the eigenstates, without referring to the explicit wavefunctions of  these states. We show that this approach offers not only a new computational methodology but also a comprehensive view of both bipartite and multipartite entanglement properties across the entire spectrum of eigenstates. Specifically, we demonstrate that states typically found in the central region of the spectrum exhibit greater multipartite entanglement, as indicated by larger QFI values, compared to states at the edges of the spectrum. In contrast, concurrence displays the opposite trend. This observed behavior is in line with the monogamy principle governing quantum entanglement.

\end{abstract}

\baselineskip 18pt
\thispagestyle{empty}
\newpage

%\begin{center}
%\textbf{\Large {Bootstrapping entanglement in quantum spin chains}}\\~\\
%\today
%\end{center}

\tableofcontents

\section{Introduction}

In the realm of many-body Hamiltonians, conventional wisdom suggests that in order to compute various properties of eigenstates—such as  correlation functions—the eigenstates themselves must first be determined. This paradigm has been questioned by the emergence of the bootstrap program, which began its life more than fifty years ago in the sphere of quantum field theory. Although initial strides were modest, the program has gained renewed vigor in the context of conformal field theory (CFT), as surveyed in \cite{Poland:2018epd,Simmons-Duffin:2016gjk}.

Broadly speaking, a technique is labeled as a ``bootstrap method'' if it aims to ascertain the properties of interest not by directly diagonalizing the Hamiltonian, but by utilizing consistency relations. This approach has been extended recently to elementary quantum mechanical systems \cite{Han:2020bkb}, demonstrating its efficacy in reexamining well-known Hamiltonians from a new angle \cite{Berenstein:2021dyf,Bhattacharya:2021btd,Aikawa:2021eai,Berenstein:2021loy,Tchoumakov:2021mnh,Du:2021hfw,Berenstein:2022unr,Nancarrow:2022wdr}. The overarching goal of these works is to enhance the methodology to outperform traditional approaches in spectrum calculation. Whether this method ultimately provides a superior way to evaluate the spectrum of quantum many-body systems remains an open question, but the foundational philosophy of the bootstrap method is certainly compelling.

In this study, our objective is to employ recent bootstrap techniques, as introduced in \cite{Han:2020bkb}, to estimate entanglement in quantum many-body systems, rather than single-body or few-body systems, without resorting to diagonalization of the Hamiltonian.
Traditionally, eigenstates are computed through either numerical or analytical means, following which ground-state entanglement measures are evaluated \cite{Amico:2007ag}. Some variations do exist; for instance, R\'enyi entanglement entropy can be calculated using swap operators \cite{Hastings:2010zka,Abanin:2012kmr,Daley:2012mov,Islam:2015ecs} without resorting to the state itself, while for two-dimensional CFT one employs the replica trick and twist operators \cite{Calabrese:2009qy}. However, these methods do not offer a holistic view of entanglement across the entire spectrum.
Our contribution in this paper is a framework capable of bootstrapping a range of properties, including energy spectrum, angular momentum, concurrence, tangle, residual tangle, and quantum Fisher information for all eigenstates or a particular sector of the eigenstates in quantum spin systems, without refereing to the explicit wavefunctions of these states.
Specifically, we focus on the well-known Lipkin-Meshkov-Glick (LMG) model \cite{Lipkin:1965tgk,Meshkov:1965ikm,Glick:1965vrc}, extending it by incorporating both longitudinal and transverse external magnetic fields.
In the bootstrap method we solve the LMG model with up to 16 sites.

Previous works have examined the bipartite entanglement properties of the LMG model's ground state \cite{Vidal:2003xgl,Vidal:2003lnl,Dusuel:2004jg}, and the quantum Fisher information has also been studied \cite{Ma:2009tsk,Salvatori:2014xlo}. Here, we go beyond these studies by evaluating both bipartite and multipartite properties of all eigenstates or a particular sector of the eigenstates.
We highlight that quantum Fisher information, as an indicator of multipartite entanglement, was initially proposed in \cite{Pezze:2007jde,Hyllus:2010zye,Toth:2010mnq} and the connection to the  dynamical susceptibilities was established in \cite{Hauke:2015bnh}.

Using the bootstrap procedure, we obtain precise values for the concurrence, tangle, residual tangle, and quantum Fisher information, which can be cross-checked using other methods.
Although the LMG model with both types of magnetic fields can be also exactly solved using angular momentum techniques, as shown in appendix~\ref{sectionLMGangmom}, the precise solution offers only modest benefits compare to the bootstrap approach if the goal is to understand the full spectrum of eigenstates. Nonetheless, we present some specific results related to concurrence of two spins and quantum Fisher information for relatively large systems of specific eigenstates in appendix~\ref{sectionLMGangmom}.
The LMG model can also be solved using exact block diagonalization, and the results obtained are presented in Appendix~\ref{sectionLMGbd}.

The remainder of this paper is structured as follows:
Section \ref{sectiontechnique} outlines the bootstrap methodology in the context of quantum spin systems.
Section \ref{sectionLMG} introduces the LMG model and identifies the operators pertinent to our bootstrap approach, offering a simple example to lay the groundwork for subsequent findings.
Section \ref{sectionconcurence} and Section \ref{sectiontangle} present calculations for concurrence, tangle, and residual tangle across the spectrum. Section \ref{sectionQFI} extends these ideas to quantum Fisher information, giving an estimation of multipartite entanglement content.
We conclude the paper with a discussion and summary in Section \ref{sectionconclusion}.
In Appendix~\ref{sectiontoy}, we demonstrate the bootstrapping process for a simple toy model.
%Appendices~\ref{sectionLMGangmom} and \ref{sectionLMGbd} present results obtained for the LMG model utilizing angular momentum and block diagonalization techniques, respectively.
In Appendix~\ref{sectionPIS}, we discuss the issue of (in)existence of the permutation invariant states in the LMG model.
In Appendix~\ref{sectionCC} we discuss the computational complexity of the bootstrap method.

\section{The bootstrap technique}\label{sectiontechnique}

For a quantum system with the Hamiltonian $H$, we want to obtain the spectrum, together with the corresponding expectation values of the operators in a set $\{ \cO_\a,\a=0,1,\cdots,\cL-1 \}$ that generates a complex vector space $\cS$, without explicitly solving the wavefunctions of the states.
We require that (1) for $\forall \cO \in \cS$ there are $H\cO\in\cS$ and $\cO H\in\cS$ and (2) for $\forall \cO,\cO' \in \cS$, there is $\cO \cO' \in \cS$.
In other words, the space $\cS$ is closed under operator multiplication.

The eigenstate $|E\rag$ with energy $E$ and the corresponding eigenstate expectation values $\lag \cO_\a \rag \equiv \lag E | \cO_\a | E \rag$ satisfy the constraints \cite{Han:2020bkb}
\bea
&& \lag [ H, \cO_\a ] \rag = 0, \label{con2} \\
&& \lag H \cO_\a \rag = E \lag \cO_\a \rag. \label{con3}
\eea
Additionally, one may require the positivity constraints on the correlation matrix
\be
\cM_{\a\b} = \lag \cO_\a^\dag \cO_\b \rag, \label{con4positive}
\ee
which is however not used in this paper.

A special case that we focus on in this paper is that $H \in \cS$, so that we can write the Hamiltonian as
\be \label{Hamiltonian}
H = \sum_\a h_\a \cO_\a,
\ee
with constant coefficients $h_\a$.
As the space $\cS$ generated by the operators $\{ \cO_\a,\a=0,1,\cdots,\cL-1\}$ is closed, we have
\be \label{tensorgabg}
\cO_\a \cO_\b = \sum_\g g_{\a\b\g} \cO_\g.
\ee
The tensor $g_{\a\b\g}$ can be calculated from%
\footnote{
Multiplying $\cO_\d$ on both sides of (\ref{tensorgabg}), we obtain
\[
\cO_\a\cO_\b\cO_\d = \sum_\g g_{\a\b\g} \cO_\g \cO_\d.
\]
Taking trace of the above equation, we get
\[
C_{\a\b\d}= \sum_\g g_{\a\b\g} B_{\g\d},
\]
which further gives (\ref{gabg}).
We have assumed that the matrix $B_{\a\b}$ is invertible, which needs to be checked for specific choice of the operator space $\cS$.
}
\be
\label{gabg}
g_{\a\b\g} = \sum_{\d} C_{\a\b\d} B^{-1}_{\d\g},
\ee
with the matrix $B_{\a\b}$ and tensor $C_{\a\b\g}$ being defined as
\bea \label{BabCabg}
&& B_{\a\b} \equiv \tr(\cO_\a\cO_\b), \nn\\
&& C_{\a\b\g} \equiv \tr(\cO_\a\cO_\b\cO_\g),
\eea
and $B^{-1}$ being the inverse matrix of $B$.

It is not guaranteed that the matrix $B_{\a\b}$ is invertible, and so one needs to choose the operator set $\cS$ properly.
With the definition $v_\a \equiv \lag \cO_\a \rag$, the constraints
(\ref{con2}) and (\ref{con3}) become
\bea
&& \sum_{\b,\g} h_\b ( g_{\b\a\g} - g_{\a\b\g} ) v_\g = 0, \label{newcon2} \\
&& \sum_{\b,\g} h_\b g_{\b\a\g} v_\g = E v_\a. \label{newcon3}
\eea
Additionally, the positive semi-definite correlation matrix (\ref{con4positive}) becomes
\be \label{positivity}
\cM_{\a\b} = \sum_{\g,\d} f_{\a\g} g_{\g\b\d}v_\d {\textrm{~is~positive~semi-definite}},
\ee
where $f$ is defined through $\cO_\a^\dag=\sum_\b f_{\a\b}\cO_\b$.
The positivity constraint (\ref{positivity}) will not be used in this paper.

When there is some symmetry generated by the operator $\cQ$, we have $[H,\cQ]=0$.
For the simultaneous eigenstates of $H$ and $\cQ$, with the eigenvalue of $Q$ for $\cQ$, we have the additional constraints
\bea
&& \lag [ \cQ, \cO_\a ] \rag = 0, \label{con4} \\
&& \lag \cQ \cO_\a \rag = Q \lag \cO_\a \rag. \label{con5}
\eea
When $\cQ\in\cS$, we may write
\be
\cQ=\sum_\a q_\a \cO_\a.
\ee
Then the constraints become
\bea
&& \sum_{\b,\g} q_\b ( g_{\b\a\g} - g_{\a\b\g} ) v_\g = 0, \label{newcon4} \\
&& \sum_{\b,\g} q_\b g_{\b\a\g} v_\g = Q v_\a. \label{newcon5}
\eea
When there are more simultaneous commuting operators, we may add more similar constraints.

The eigenvalues of symmetry generators can help distinguish degenerate states in the spectrum. However, if two states in the spectrum have the same energy and all the operators in the space $\cS$ have the same expectation values in the two states, the bootstrap technique mentioned above would not be able to distinguish between the two states.

The goal of the bootstrap procedure in this paper is to solve the constraint equations (\ref{newcon2}) and (\ref{newcon3}), possibly with the additional constraints (\ref{newcon4}) and (\ref{newcon5}), and obtain the spectrum of the Hamiltonian and the corresponding expectation values.
Given that these constraints are expressed as exact linear equations, the energy and the corresponding expectation values will be precise whenever a solution exists.
The key aspect is to choose the space $\cS$ generated by an appropriate set of operators.
Typically, all the possible eigenvalues $Q$ of a symmetry generator $\cQ$ are known, and the spectrum of the Hamiltonian can be classified into different sectors based on the value of $Q$.
One may choose to use the constraints (\ref{newcon4}) and (\ref{newcon5}) with a specific value $Q$ to obtain the spectrum and the corresponding expectation values in a specific sector of $Q$.

\section{Bootstrapping LMG model}\label{sectionLMG}

We consider the LMG model with the Hamiltonian \cite{Lipkin:1965tgk,Meshkov:1965ikm,Glick:1965vrc}
\be \label{HLMG}
H = - \f{1}{N} \sum_{1\leq j_1 < j_2 \leq N} \Big( \f{1+\k}{4} \s_{j_1}^x \s_{j_2}^x + \f{1-\k}{4} \s_{j_1}^y \s_{j_2}^y \Big)
    - \f12 \sum_{j=1}^N ( \l_x \s_j^x + \l_z \s_j^z ),
\ee
with both a longitudinal field $\l_x$ and a transverse field $\l_z$.
Here the integer number $N$ is the total number of sites in the system.
We define
\be \label{JxJyJz}
J_x \equiv \f12 \sum_{j=1}^N \s_j^x, ~~
J_y \equiv \f12 \sum_{j=1}^N \s_j^y, ~~
J_z \equiv \f12 \sum_{j=1}^N \s_j^z,
\ee
which satisfy the standard commutation relations of the components of the angular momentum operator.
The Hamiltonian can be recast as
\be \label{HLMGitoJxJyJz}
H = -\f1N \Big( \f{1+\k}{2} J_x^2 + \f{1-\k}{2} J_y^2 \Big) +\f14 - \l_x J_x - \l_z J_z,
\ee
in terms of which the LMG model can be solved exactly.
We show the solutions of the LMG model from angular momentum method in appendix~\ref{sectionLMGangmom} and the solutions from block diagonalization in appendix~\ref{sectionLMGbd}.

We choose the complex linear space $\cS$ generated by the operators
\be \label{cOabc}
\{ \cO_{abc} \equiv J_x^a J_y^b J_z^c, 0 \leq a+b+c \leq N \}.
\ee
In the space $\cS$ the number of linearly independent operators are
\be
\cL = \f{(N+1)(N+2)(N+3)}{6},
\ee
including the identity operator.

For the LMG model, the operator $J^2=J_x^2+J_y^2+J_z^2$ is a conserved quantity, and we can classify the spectrum of the LMG model into several sectors by using angular quantum numbers $l=l_{\rm min},l_{\rm min}+1,\cdots,\f{N}{2}$, where $l_{\rm min}=0$ for even integer $N$ and $l_{\rm min}=\f12$ when $N$ is an odd integer.
Note that $\lag J^2 \rag=l(l+1)$.
We write the expectation values as $v_{abc} \equiv \lag \cO_{abc} \rag$.
The constraints (\ref{newcon2}), (\ref{newcon3}), (\ref{newcon4}) and (\ref{newcon5}) are linear equations for the expectation values $v_{abc}$.
The strategy is to eliminate as many as possible the expectation values $v_{abc}$ from these linear equations.
If all the parameters $v_{abc}$ are eliminated, we would obtain the equations of $E$, from which the exact values of $E$ are obtained.
Then we substitute each exact value of $E$ in the original linear equations and get the corresponding expectation values.
If only part of the parameters $v_{abc}$ are eliminated, we would obtain the equations of $E$ and the remaining parameters $v_{abc}$.
Then we need to use the positivity constraint (\ref{positivity}) to restrict the energy and the remaining parameters $v_{abc}$ in some regions.
It turns out that for all the examples we consider in this paper, we obtain the exact values of the energy and the corresponding expectation values without using the positivity constraint (\ref{positivity}).

We take $\k=\l_x=\l_z=1$ and $N=2$ as an example.
The sector $l=0$ is easy to solve.
We have the constraints (\ref{newcon2}) and (\ref{newcon3}), as well as (\ref{newcon4}) and (\ref{newcon5}) with the conserved quantity $\cQ=J^2$ and its eigenvalue $Q=0$. These are linear equations of 9 parameters $v_{200}$, $v_{020}$, $v_{002}$, $v_{110}$, $v_{101}$, $v_{011}$, $v_{100}$, $v_{101}$ and $v_{001}$, and the coefficients of these linear equations may depend on $E$.
The constraints (\ref{newcon4}) are trivial, and the constraints (\ref{newcon5}) give
\be
v_{abc}=0.
\ee
Substitute $v_{abc}=0$ in the constraints (\ref{newcon2}) and (\ref{newcon3}), we get the constraint for the energy
\be
4 E-1=0,
\ee
which gives the solution $E=0.25$.

The sector $l=1$ is more typical to show the algorithm.
We have the constraints (\ref{newcon2}) and (\ref{newcon3}), as well as (\ref{newcon4}) and (\ref{newcon5}) with $\cQ=J^2$ and $Q=2$.
There are many constraint equations, and we just give some examples
\bea
&& v_{001} + 2 \ii v_{110} =0, ~~ \ii v_{101}=0, \nn\\
&& \f14 - E - v_{001} - \f12 v_{200}=0, \nn\\
&& v_{200}+v_{020}+v_{002}=2.
\eea
We first eliminate the 9 parameters in the constraints.
In other words, we get the 9 parameters in terms of yet unknown energy $E$
\bea \label{vabcsolutions}
&& v_{200} = \frac{48 E^2-32 E+309}{568}, ~~
   v_{020} = -\frac{496 E^2+48 E-2203}{2272}, \nn\\
&& v_{002} = \frac{304 E^2+176 E+1105}{2272}, ~~
   v_{110} = -\frac{\ii (112 E^2+304 E-131)}{1136},  \nn\\
&& v_{101} = \frac{400 E^2+112 E-549}{1136},  ~~
   v_{011} = \frac{\ii (176 E^2-496 E-287)}{2272}, \nn\\
&& v_{100} = \frac{176 E^2-496 E-287}{1136},  ~~
   v_{010} = 0,  \nn\\
&& v_{001} =-\frac{112 E^2+304 E-131}{568}.
\eea
%In the process, we need to avoid putting any polynomial of the energy $E$ in the denominator as the polynomial may be zero.
In the constraints, we are left with the equation for the energy
\be
64 E^3+16 E^2-132 E-1=0,
\ee
which has the solutions
\be
E=\{-1.563,-0.007569,1.321\}.
\ee
Putting the exact values of $E$ back in (\ref{vabcsolutions}), we get the corresponding exact expectation values
\bea
&& v_{200} = \{0.8386,0.5444,0.6170\}, ~~
   v_{020} = \{0.4693,0.9698,0.5610\}, \nn \\
&& v_{002} = \{0.6922,0.4858,0.8220\}, ~~
   v_{110} = \{0.2927 \ii,0.1173 \ii,-0.4101 \ii\}, \nn \\
&& v_{101} = \{0.2229,-0.4840,0.2611\},  ~~
   v_{011} = \{0.4042 \ii,-0.1247 \ii,-0.2795 \ii\}, \nn \\
&& v_{100} = \{0.8084,-0.2493,-0.5590\},  ~~
   v_{010} = \{0,0,0\},  \nn\\
&& v_{001} = \{0.5854,0.2347,-0.8201\}.
\eea

The calculations for larger $N$ are similar and we solve the LMG model (\ref{HLMG}) up to $N=16$.
Generally, in the sector with quantum number $l$, the constraint equation of the energy $E$ is a polynomial of order $2l+1$.
For example, when $\k=\l_x=\l_z=1,N=16,l=1$ we have
\be
1024 E^3-640 E^2-1916 E+439 = 0,
\ee
and when $\k=\l_x=\l_z=1,N=16,l=3$ we have
\be
262144 E^7-7383040 E^5+690944 E^4+51517252 E^3-6057569 E^2-74528784 E+4584836=0.
\ee

From the bootstrap approach, we obtain the spectrum and the corresponding expectation values of a chosen set of operators without having explicit knowledge of the wavefunction of the system. Subsequently, we will use these expectation values to determine the entanglement properties of the states.

\section{Concurrence}\label{sectionconcurence}

In a mixed state, it is typically difficult to quantify the degree of entanglement between two subsystems. However, for two qubits, the amount of entanglement can be easily determined using a measure called concurrence.
For two qubits in state with density matrix $\r$, the concurrence is defined as \cite{Wootters:1997id}
\be \label{Cmaxl1ml2ml3ml40}
C = \max \{ \z_1 -\z_2 -\z_3 -\z_4 , 0 \},
\ee
where $\{\z_1,\z_2,\z_3,\z_4\}$ in descending order $\z_1 \geq \z_2 \geq \z_3 \geq \z_4$ are the eigenvalues of the positive semidefinite operator
\be
R =  \sr {\sr\r \s_1^y\s_2^y \r^* \s_1^y\s_2^y \sr \r} .
\ee
Note that the complex conjugate of the two-qubit density matrix $\r$ is taken in the eigenbasis of $\s_1^z\s_2^z$.
A general two-qubit density matrix takes the form
\be \label{rhotwoqubit}
\r = \sum_{\a,\b=0}^3 f_{\a\b} \s_1^\a \s_2^\b, ~~ \s^\a=\{ 1, \s^x, \s^y, \s^z \},
\ee
with the parameters determined by the double-spin correlation functions
\be \label{coefficients}
f_{\a\b} = \f14 \tr(\r \s_1^\a \s_2^\b).
\ee
It should be noted that $f_{00}=\f14$ always holds. A general two-qubit state is fully described by 15 real parameters, from which one can determine the density matrix and calculate the concurrence using (\ref{Cmaxl1ml2ml3ml40}). However, if more information about the symmetry of state is known, we do not actually need all the 15 real parameters to calculate the concurrence \cite{Amico:2003tdv,Fubini:2005ber,Amico:2007ag}.

The LMG model has the permutation symmetry $S_N$. In other words, the Hamiltonian (\ref{HLMG}) is invariant under the exchange of two arbitrary sites.
However, as the permutation symmetry $S_N$ is non-Abelian, it is not guaranteed that all the eigenstates can be written in permutation invariant forms.
In appendix~\ref{sectionPIS}, we show that all the permutation invariant states are in the sector $l=N/2$.
In a permutation invariant state, the concurrence between two arbitrary sites is the same.
In the present setup of the bootstrap, we cannot calculate the concurrence for energy eigenstates with $l<\f{N}{2}$.

\begin{figure}[t]
  \centering
  \includegraphics[width=0.98\textwidth]{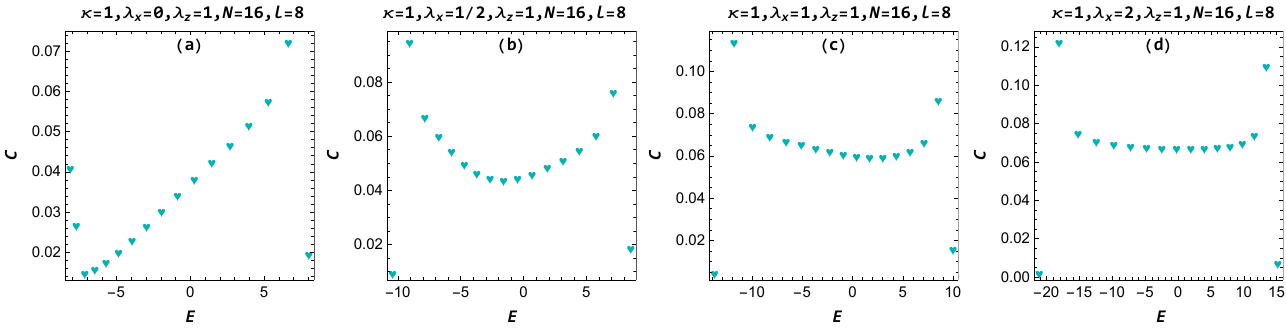}
  \caption{The concurrence between two qubits in the $l=N/2$ sector of the LMG model calculated using the bootstrap method.
  In different columns we have chosen different values for the longitudinal and transverse fields $\l_x$ and $\l_z$.}
  \label{FigureConcurrence}
\end{figure}

From the bootstrap, we obtain the energy spectrum and the expectation values $\lag J_x \rag$, $\lag J_y \rag$, $\lag J_z \rag$, $\lag J_x^2 \rag$, $\lag J_y^2 \rag$, $\lag J_z^2 \rag$, $\lag J_x J_y \rag$, $\lag J_x J_z \rag$, $\lag J_y J_z \rag$.
For the coefficients (\ref{coefficients}), there is always $f_{00}=\f14$.
In a permutation invariant state, i.e. a state in the sector $l=N/2$, there is the invariance of the two-qubit density matrix $\rho$ under the exchange of the two qubits, which means that $f_{\a\b}=f_{\b\a}$.
This leaves us with 9 independent parameters to describe the state
\bea \label{coefficientsf01tof33}
&& f_{01}=\f{1}{2N} \lag J_x \rag, ~~
   f_{02}=\f{1}{2N} \lag J_y \rag, ~~
   f_{03}=\f{1}{2N} \lag J_z \rag, \nn \\
&& f_{11}=\f{1}{N-1} \Big( \f1N \lag J_x^2 \rag - \f14 \Big), ~~
   f_{22}=\f{1}{N-1} \Big( \f1N \lag J_y^2 \rag - \f14 \Big),\nn \\
&& f_{33}=\f{1}{N-1} \Big( \f1N \lag J_z^2 \rag - \f14 \Big), ~~
   f_{12}=\f{1}{N(N-1)} \Big( \lag J_x J_y \rag - \f\ii2 \lag J_z \rag \Big), \nn\\
&& f_{13}=\f{1}{N(N-1)} \Big( \lag J_x J_z \rag + \f\ii2 \lag J_y \rag \Big), ~~
   f_{23}=\f{1}{N(N-1)} \Big( \lag J_y J_z \rag - \f\ii2 \lag J_x \rag \Big).
\eea
Note that these coefficients (\ref{coefficientsf01tof33}) are only valid for permutation invariant states, i.e. states in the sector $l=N/2$.
Then we construct the density matrix (\ref{rhotwoqubit}) and calculate the concurrence.
We plot the results in figure~\ref{FigureConcurrence}.

Due to the specific operator space generator (\ref{cOabc}) that we have chosen, the calculations of concurrence are limited to permutation-invariant states. To extend to more general states, it is necessary to consider alternative selections for the operator space $\cS$.

\section{Tangle and residual tangle}\label{sectiontangle}

When an entire system is in a pure state, one can quantify the entanglement between a single qubit and the remainder of the system using the entanglement entropy, i.e., the von Neumann entropy of the one-qubit density matrix $\rho$ given by
\be
S = - \tr (\r\log\r).
\ee
Alternatively, one may use the tangle
\be \label{teq4detr}
\t = 4 \det \r.
\ee
The entanglement entropy and tangle are equivalent and they are simply related as
\be \label{Seqft}
S = - \f{1+\sqrt{1-\t}}{2} \log \f{1+\sqrt{1-\t}}{2} - \f{1-\sqrt{1-\t}}{2} \log \f{1-\sqrt{1-\t}}{2}.
\ee
Note that both formulas (\ref{teq4detr}) and (\ref{Seqft}) only apply to one-qubit, and they represent the entanglement between the single qubit and the rest $N-1$ qubits.

The tangle between two qubits is $\t=C^2$ with $C$ being the concurrence, which applies even when the two qubits are in a mixed state.
The monogamy of the entanglement is encoded in the Coffman-Kundu-Wootters (CKW) inequality \cite{Coffman:1999jd,Osborne:2005mqz}
\be
\sum_{j=2}^N C_{1j}^2 \leq \t_{1(23\cdots N)},
\ee
where $C_{1j}$ is the concurrence between the first qubit and the $j$-th qubit and $\t_{1(23\cdots N)}$ is the tangle between the first qubit and the rest qubits.
From the CKW inequality one may define the residual tangle as \cite{Coffman:1999jd}
\be
\D \t_{1(23\cdots N)} = \t_{1(23\cdots N)} - \sum_{j=2}^N C_{1j}^2.
\ee
As $\t_{1(23\cdots N)}$ measures the entanglement between the first qubit and the rest qubits and $C_{1j}^2$ measures the entanglement between the first qubit and the $j$-th qubit, the residual tangle encodes the part of the multipartite entanglement that is not encoded in double-qubit entanglement.

%In LMG model, we consider the energy eigenstates that are invariant under the exchange of two arbitrary qubits.
In any translation invariant eigenstate of the LMG model, the single-qubit density matrix is written as
\be \label{rhoonequbit}
\r = \sum_{\a=0}^3 f_\a \s^\a,
\ee
with $f_0=\f12$ and
\be \label{f1f2f3}
f_1=\f{1}{N}\lag J_x \rag, ~~ f_2=\f{1}{N}\lag J_y \rag, ~~ f_3=\f{1}{N}\lag J_z \rag.
\ee
%which apply for any translation invariant states.
The tangle between one qubit and the rest qubits is
\be \label{tangle}
\t=1-\f{4}{N^2} ( \lag J_x \rag^2 + \lag J_y \rag^2 + \lag J_z \rag^2 ).
\ee
For states that are not only translation invariant but also permutation invariant, the CKW inequality is
\be
(N-1) C^2 \leq \t,
\ee
with $C$ being the concurrence between two arbitrary qubits.
The residual tangle is
\be \label{residualtangle}
\D\t=\t-(N-1)C^2.
\ee
We emphasize that the tangle (\ref{tangle}) applies to any translation invariant states, while the residual tangle (\ref{residualtangle}) only applies to permutation invariant states.

\begin{figure}[t]
  \centering
  \includegraphics[width=0.98\textwidth]{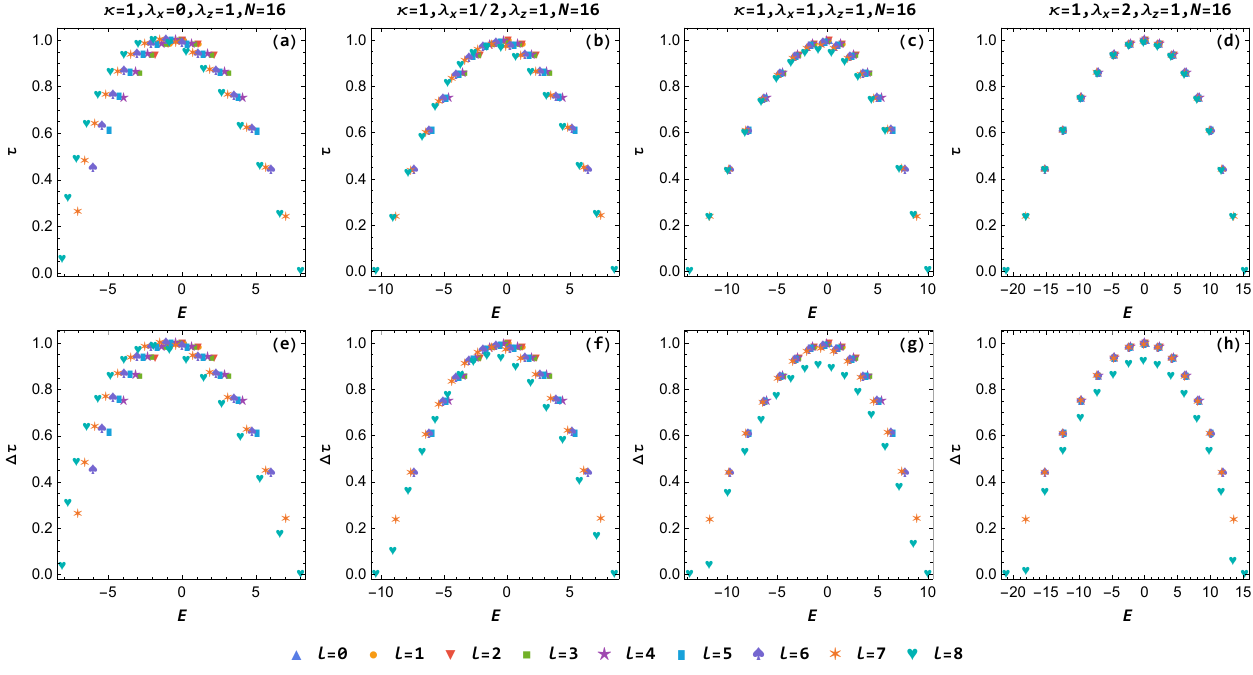}
  \caption{The tangle (top panels) and residual tangle (bottom panels) between a single qubit and the rest qubits in the LMG model obtained from the bootstrap analysis. We use different colored symbols to represent states with different angular quantum number $l$.
  In different columns we have chosen different values for the longitudinal and transverse fields $\l_x$ and $\l_z$.}
  \label{FigureTangle}
\end{figure}

The tangle for states in all the sectors $l=0,1,\cdots,N/2$ and residual tangle in the sector $l=N/2$ are obtained from the bootstrap analysis, as shown in figure~\ref{FigureTangle}.
%The tangle is approximately a function of the energy $E$ and independent of the quantum number $l$.
The residual tangle is almost the same as the tangle. %except that the residual tangle in states with quantum number $l=\f{N}{2}$ is slightly smaller than the tangle.
States in the middle of the spectrum tend to have larger tangles and residual tangles, while states near the ground state and the highest-energy state have smaller tangles and residual tangles.
This indicates that states in the middle of the spectrum have stronger entanglement and that most of the entanglement therein is stored as multipartite entanglement.

In section~\ref{sectionconcurence} for a permutation invariant state we have constructed the two-qubit density matrix (\ref{rhotwoqubit}) from the 9 real independent parameters (\ref{coefficientsf01tof33}), and in this section we have constructed the one-qubit density matrix (\ref{rhoonequbit}) from the 3 real independent parameters (\ref{f1f2f3}).
Similarly, a general $n$-qubit density matrix can be written as
\be
\r = \sum_{\a_1,\a_2,\cdots,\a_n=0}^3 f_{\a_1\a_2\cdots\a_n}\s_1^{\a_1}\s_2^{\a_2}\cdots\s_n^{\a_n},
\ee
with
\be
f_{\a_1\a_2\cdots\a_n} \equiv \f{1}{2^n} \tr( \r \s_1^{\a_1}\s_2^{\a_2}\cdots\s_n^{\a_n} ).
\ee
For a general density matrix, the number of independent parameters is $4^n-1$.
Given the permutation invariance of the state, the number of real independent parameters become $\f{(n+1)(n+2)(n+3)}{6}-1$, and these parameters can be obtained from the $\f{(n+1)(n+2)(n+3)}{6}-1$ expectation values
\be
\lag J_x^a J_y^b J_z^c \rag, ~ 1 \leq a+b+c \leq n.
\ee
With the explicit $n$-qubit density matrix, one can calculate any quantity defined from the density matrix, including the entanglement entropy and other entanglement measures.

\section{Quantum Fisher information}\label{sectionQFI}

From the bootstrap analysis, we can also calculate the quantum Fisher information (QFI), which can be used as an entanglement witness to characterize multipartite entanglement as shown in \cite{Pezze:2007jde,Hyllus:2010zye,Toth:2010mnq}.

The QFI quantifies the sensitivity of a quantum state to a parameter. For a general pure state \(|\psi\rangle\) and a Hermitian operator \(\mathcal{A}\), the QFI characterizes the sensitivity of the state \(e^{-i\mathcal{A}\theta}|\psi\rangle\) to the parameter \(\theta\), which corresponds to four times the variance of \(\mathcal{A}\) in the state \(|\psi\rangle\)
\be
F_{\mathcal{A}} = 4 \left( \langle \psi | \mathcal{A}^2 | \psi \rangle - \langle \psi | \mathcal{A} | \psi \rangle^2 \right).
\ee
In a pure state, for the operators $J_{x,y,z}$ there are the QFI
\be
F_{x,y,z} = 4 ( \lag J_{x,y,z}^2 \rag - \lag J_{x,y,z} \rag^2  ).
\ee
For a general direction
\be
r = ( \sin\th\cos\phi, \sin\th\sin\phi, \cos\th  ),
\ee
there is the operator $J_r = r_x J_x + r_y J_y + r_y J_y$, and the corresponding QFI is
\be
F_r = 4 ( \lag J_r^2 \rag - \lag J_r \rag^2  ).
\ee
For a state one may define the QFI related quantities
\be
F_{\rm max} \equiv \max_{\th,\phi} F_r,
\ee
and
\be
F_{\rm sum} \equiv \sum_{i=x,y,z} F_i.
\ee
A pure state of $m$ particles is called separable if it can be written as the direct product of $m$ one-particle states.
For all separable pure states, there are \cite{Pezze:2007jde}
\be \label{FrmmaxleqL}
F_{\rm max} \leq N
\ee
and \cite{Toth:2010mnq}
\be \label{Frmsumleq2L}
F_{\rm sum} \leq 2 N.
\ee
If for a pure state there is
\be \label{FrmmaxgL}
F_{\rm max} >N,
\ee
or
\be \label{Frmsumg2L}
F_{\rm sum} >2N,
\ee
there exists entanglement in the state.
In other words, for a pure state of $N$ qubits we have
\be
({\rm not~entangled}) \Leftrightarrow {\rm separable} \Rightarrow (F_{\rm max} \leq N {\rm ~and~} F_{\rm sum} \leq 2 N),
\ee
or equivalently
\be
{\rm entangled} \Leftrightarrow ({\rm not~separable}) \Leftarrow (F_{\rm max} > N {\rm ~or~} F_{\rm sum} > 2 N).
\ee%
If $F_{\rm max} \leq N$ and $F_{\rm sum} \leq 2N$, it does not necessarily mean there is no entanglement.
The constraints (\ref{FrmmaxgL}) and (\ref{Frmsumg2L}) are just sufficient but not necessary conditions for the existence of entanglement.

A pure state is called $k$-producible if it can be written as the direct product of several pure states, with each pure state involving at most $k$ particles. It is easy to see from the definition that a $k$-producible state is automatically $(k+1)$-producible. The special case of a 1-producible state is simply a separable state.
A $k$-producible state that is not $(k-1)$-producible is genuinely $k$-partite entangled.
For general $k$-producible states, there are upper bounds \cite{Toth:2010mnq}
\be
F_{\rm max} \leq F_{\rm max}^k,
\ee
\be
F_{\rm sum} \leq F_{\rm sum}^k,
\ee
with the definitions
\be \label{Frmmaxk}
F_{\rm max}^k \equiv n k^2+(N-nk)^2,
\ee
\be
F_{\rm sum}^k \equiv \lt\{
\ba{ll}
nk(k+2)+(N-nk)(N-nk+2) & {\rm if~} N-nk \neq 1 \\
nk(k+2)+2              & {\rm if~} N-nk=1
\ea
\rt. \!\!\!\! ,
\ee
and $n$ being the integer part of $\f{N}{k}$.
When $k=1$, the state is just separable, and we have $F_{\rm max}^1=N$, which is the same as the bound (\ref{FrmmaxleqL}), and $F_{\rm sum}^1=3N$, which is looser than the bound (\ref{Frmsumleq2L}).
In the following we will redefine for the $k=1$ case $F_{\rm sum}^1 \equiv 2N$.
In other words, we have
\be \label{Frmsumk}
F_{\rm sum}^k \equiv \lt\{
\ba{lll}
2N                     & {\rm if~} k=1 \\
nk(k+2)+(N-nk)(N-nk+2) & {\rm if~} k > 1 {\rm ~and~} N-nk \neq 1 \\
nk(k+2)+2              & {\rm if~} k > 1 {\rm ~and~} N-nk=1
\ea
\rt. \!\!\!\! .
\ee
Note that a $k$-producible state is automatically a $(k+1)$-producible state, and there are always $F_{\rm max}^k \leq F_{\rm max}^{k+1}$ and $F_{\rm sum}^k \leq F_{\rm sum}^{k+1}$ as required.

When for a state there is
\be
F_{\rm max} > F_{\rm max}^k,
\ee
or
\be
F_{\rm sum} > F_{\rm sum}^k,
\ee
the state has at least genuine multipartite entanglement of $k+1$ particle, or, in other words, the entanglement depth of the state is larger than $k$.
Especially, when
\be
F_{\rm max} > F_{\rm max}^{N-1}=(N-1)^2+1,
\ee
or
\be
F_{\rm sum} > F_{\rm sum}^{N-1}=N^2+1,
\ee
the entanglement depth of the state takes the largest value $N$, and any tripartite subsystems in the system is genuinely multipartite entangled.

\begin{figure}[t]
  \centering
  \includegraphics[width=0.98\textwidth]{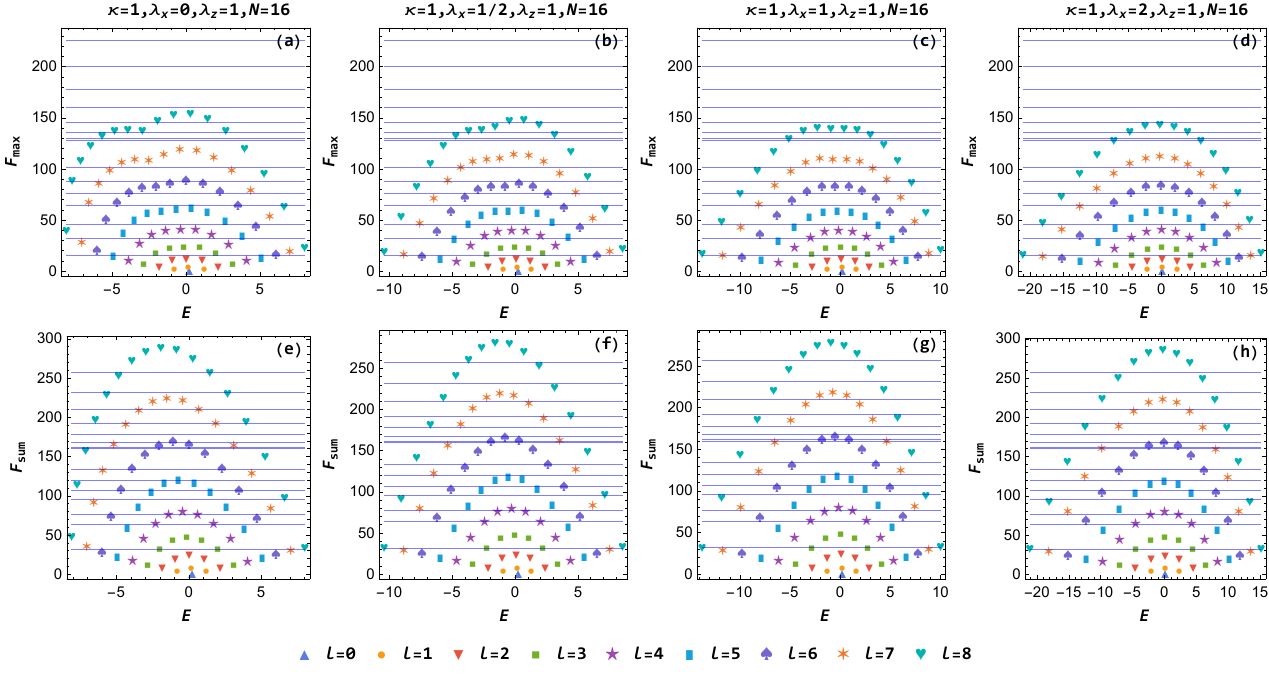}
  \caption{Quantum Fisher information (QFI) in the spectrum of the LMG model from bootstrap. We use different colored symbols to represent states with different angular quantum number $l$. The thin blue lines are the bounds (\ref{Frmmaxk}) and (\ref{Frmsumk}), with $k$ taking values in the range $[1,N-1]$ from the bottom to the top.
  In different columns we have chosen different values for the longitudinal and transverse fields $\l_x$ and $\l_z$.}
  \label{FigureQFI}
\end{figure}

From the bootstrap analysis, we obtain the energy spectrum and the corresponding expectation values, as well as the quantum Fisher information (QFI). We show the results for various parameters in figure~\ref{FigureQFI}. For small $l$, the bounds (\ref{Frmmaxk}) and (\ref{Frmsumk}) give approximately the same results, while for large $l$, the bound (\ref{Frmsumk}) is more effective than (\ref{Frmmaxk}). We observe that states in the middle of the spectrum and with larger quantum number $l$ tend to have stronger multipartite entanglement. For every energy window, the multipartite entanglement increases with $l$.
The results of QFI in figure~\ref{FigureQFI} are in agreement with the results of tangle and residual tangle in figure~\ref{FigureTangle}.
Entanglement encodes a type of quantum correlation between qubits at different sites. The results of QFI show that in a state at the edge of the spectrum, the qubits tend to be more weakly correlated, while in a state in the middle of the spectrum, the qubits tend to be more strongly correlated.
For states in the same energy window, a state with a large quantum number $l$ also exhibits stronger qubit correlation.

\section{Conclusion and discussions}\label{sectionconclusion}

In this study, we have elaborated on a methodology for determining the entanglement attributes of eigenstates for particular Hamiltonians, all without having to directly compute the eigenstates. The core approach employs consistency relations, as given by Eqs. (\ref{newcon2}) and (\ref{newcon3}), augmented with additional constraints as in Eqs. (\ref{newcon4}) and (\ref{newcon5}).
Utilizing these, we can derive the energies and operator expectation values for all eigenstates in the model with up to 16 sites.
From these expectation values, entanglement measures such as concurrence between two qubits, tangle, residual tangle, and quantum Fisher information can be explicitly calculated for all eigenstates or a particular sector of the eigenstates simultaneously.
The results from bootstrap are consistent with those obtained from the angular momentum and block diagonalization in appendices~\ref{sectionLMGangmom} and \ref{sectionLMGbd}.
Perhaps unsurprisingly, our findings reveal that eigenstates in the central part of the spectrum for the LMG model exhibit significant multipartite entanglement but limited concurrence. This observation aligns well with the expected behavior based on the monogamy property of entanglement. We have corroborated this through calculations involving both residual tangle and quantum Fisher information.

While the current paper is focused on the Lipkin-Meshkov-Glick (LMG) model, which includes both transverse and longitudinal external magnetic fields, the methodology is extendable to Hamiltonians of the form given in Eq. (\ref{Hamiltonian}) without any major modifications.
The central aspect of this method lies in selecting an appropriate set of operators (\ref{cOabc}), which are applicable to all Hamiltonians of the format (\ref{Hamiltonian}) and yield the desired outcomes of concurrence, tangle, and QFI. However, for Hamiltonians with more generalized forms, alternative operator configurations must be considered.

It's worth noting that the computational complexity of our bootstrap method increases rapidly with the size of the system, which we discuss in details in appendix~\ref{sectionCC}.
However, this shouldn't be viewed as a limitation since it allows us to obtain all correlation functions for all eigenstates in a unified manner. One potential avenue for future research can be identifying an effective truncation scheme that enables the extraction of ground-state or low-energy-state properties with less computational effort. The specific nature of such a truncation remains an open question at this stage.

Additionally, due to the Hamiltonian's permutation property as expressed in Eq. (\ref{Hamiltonian}), we did not need to employ the positivity condition shown in Eq. (\ref{positivity}). For short-range Hamiltonians, such as the transverse field Ising chain with or without longitudinal magnetic fields, additional operators would need to be incorporated into the bootstrap framework to calculate entanglement measures like the concurrence of two qubits. Efficient strategies for identifying the necessary operators for these kinds of models require further investigation.

\section*{Acknowledgements}

We thank Marcello Dalmonte for helpful discussions, comments and encouragement.
We also thank Bin Chen and Markus Heyl for helpful discussions.
J.Z.\ acknowledges support from the National Natural Science Foundation of China (NSFC) through grant number 12205217.
A.J.\ acknowledges support from a research fellowship from the The Royal Commission for the Exhibition of 1851.
M.A.R.\ thanks CNPq and FAPERJ (grant number 210.354/2018) for partial support.

\appendix

\section{Bootstrapping a toy model} \label{sectiontoy}

We consider the simple toy model with Hamiltonian
\be
H = \f12 \sum_{j=1}^N \s_j^z.
\ee
We define $J_z \equiv \f12 \sum_{j=1}^N \s_j^z$ and choose the set $\cS$ generated by the operators
\be
\{ J_z^\a, \a = 0,1,2,\cdots,N \}.
\ee
We only consider the cases with even integer $N$, and it is similar for cases with odd integer $N$.
The operator $J_z$ has eigenvalues $0,\pm1,\cdots,\pm\f{N}{2}$, and from the Cayley-Hamilton theorem
\be
J_z \prod_{m=1}^{N/2} ( J_z^2 - m^2 ) = 0,
\ee
one may write higher order powers of $J_z$ in terms of operators in $\cS$.
For example, there are
\bea
&& N=2: ~ J_z^3 = J_z, \nn\\
&& N=4: ~ J_z^5 = 5 J_z^3 - 4 J_z. %,\nn\\
%&& N=6: ~ J_z^7 = 14 J_z^3 - 49 J_z^3 + 36 J_z.
\eea

For an energy eigenstate $|E\rag$, we define $v_\a \equiv \lag E| J_z^\a |E\rag$.
Note that $J_z^0=I$ is just the identity operator that so there is $v_0=1$.
From the constraint (\ref{newcon2}) with $\a=0,1,\cdots,N-1$, we get
\be
v_{\a+1} = E v_\a, ~ \a=0,1,\cdots,N-1,
\ee
and so we have
\be
v_\a = E^\a, \a=0,1,\cdots,N.
\ee
From the constraint (\ref{newcon2}) with $\a=N$, we further obtain
\be
E \sum_{m=0}^{N/2}( E^2 - m^2 ) = 0.
\ee
Finally we get $N+1$ different solutions of the energy and expectation values
\be
E = m, ~
v_\a = m^\a, ~ \a = 0,1,\cdots,N-1, ~ m =0, \pm 1, \cdots, \pm \f{N}{2},
\ee
which are just the exact solutions.
Note that to obtain the above solutions we have not used the constraint (\ref{newcon3}) which is automatically satisfied by the solutions to the constraints (\ref{newcon2}).

Needless to say, in this appendix, we have used a difficult method to solve a simple problem.

\section{Solving LMG model from algebra of angular momentum} \label{sectionLMGangmom}

The spectrum of the LMG model can be  solved exactly using the algebra of angular momentum, as originally proposed by Lipkin in \cite{Lipkin:1965tgk}. This approach involves writing the Hamiltonian in a different basis, which results in a significantly simpler form.

At each site, there exists the fundamental representation $\bf 2$, i.e. the spin-1/2 representation, of the su(2) algebra. For the entire system, the representation of the su(2) algebra is reducible and can be expressed as
\be
{\bf 2}^{\otimes N} = \bigoplus_{l=l_{\rm min}}^{N/2} ({\bf 2\ibl + 1})_{d_l},
\ee
where the duplicate number of the $({\bf 2\ibl+1})$ representation, i.e. the spin-$l$ representation, is given by
\be \label{dl}
d_{l} = \frac{2l+1}{N+1}C_{N+1}^{\frac{N}{2}+l+1}.
\ee
There are $l_{\rm min}=0$ for an even integer $N$, and $l_{\rm min}=\frac{1}{2}$ for an odd integer $N$.
For the special case where $l=N/2$, the duplicate number $d_{N/2}=1$.

Noting the Hamiltonian (\ref{HLMGitoJxJyJz}), for $l=0,\f12,1,\f32,\cdots$ we define the operator
\be
H^{(l)}  = -\f1N \Big[ \f{1+\g}{2} ( J_x^{(l)} )^2 + \f{1-\g}{2} ( J_y^{(l)} )^2 \Big] +\f14 I_{2l+1} - \l_x J_x^{(l)} -\l_z J_z^{(l)},
\ee
with $J_x^{(l)}$, $J_y^{(l)}$, $J_z^{(l)}$ being the components of the spin-$l$ angular momentum and $I_{2l+1}$ being the $(2l+1) \times (2l+1)$ identity matrix.
Then from representation reduction, the Hamiltonian (\ref{HLMG}) is similar to
\be \label{Hsim}
H \sim \bigoplus_{l=l_{\rm min}}^{N/2} H^{(l)}_{d_{l}}.
\ee
We have used the same symbol $d_l$ (\ref{dl}) as the duplicate number of each block $H^{(l)}$.
For the Hamiltonian $H$, every eigenvalue of $H^{(l)}$ has degeneracy $d_l$.

As shown in (\ref{Hsim}), the Hamiltonian can be decomposed into direct product of several blocks in a specific basis, and the explicit transformation relation between the block-diagonal basis and the $\sigma^z$ basis is not of concern to us. In each block, one may obtain the eigenvalues and eigenstates of the Hamiltonian and compute the expectation values of operators that can be expressed in terms of $J_x, J_y, J_z$ given in (\ref{JxJyJz}). For the sector $l=N/2$, the energy eigenstates are superpositions of the $N+1$ states.
\be \label{statesL2m}
| N/2 , m \rag = \f{1}{\sqrt{C_{N/2}^m}}
\sum_{1\leq j_1 < j_2 < \cdots < j_m \leq N}
| \ua_{j_1} \ua_{j_2} \cdots \ua_{j_m}\rag, ~
m=\f{N}{2},\f{N}{2}-1,\cdots,-\f{N}{2}.
\ee
Here, $| \uparrow_{j_1} \uparrow_{j_2} \cdots \uparrow_{j_m} \rangle$ denotes the state in which the $m$ sites $j_1, j_2, \cdots, j_m$ have up spins and all other sites have down spins.
All the states (\ref{statesL2m}) are invariant under an arbitrary permutation of the sites, and so all the eigenstates in the sector $l = N/2$ are also permutation invariant.
As we show in appendix~\ref{sectionPIS}, the eigenstates in the other sectors are not permutation invariant.

\begin{figure}[p]
  \centering
  \includegraphics[width=0.98\textwidth]{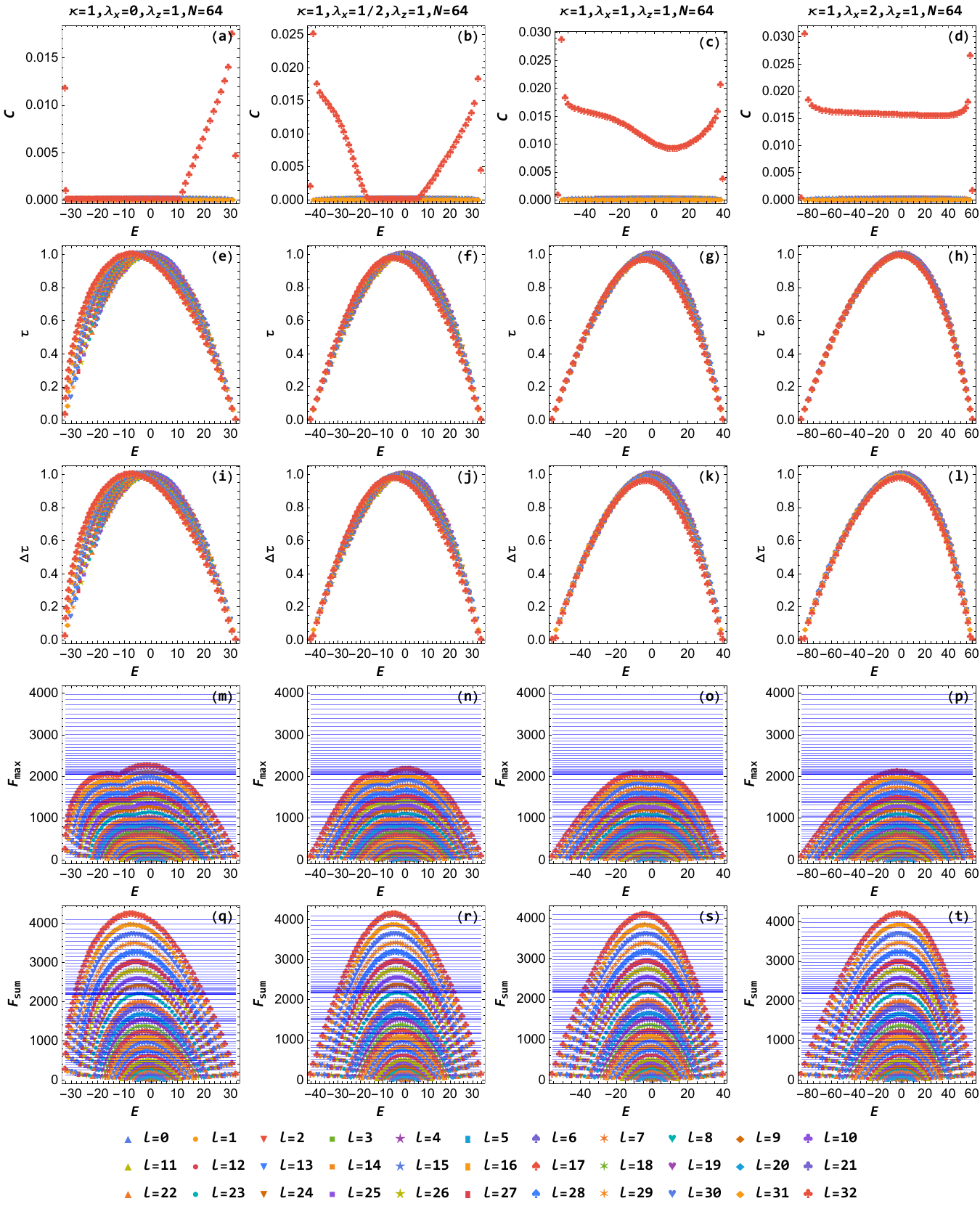}
  \caption{The concurrence between two sites (the first, second and third rows), tangle (fourth row) and residual tangle (fifth row) between one site and the other sites, as well as quantum Fisher information (last two rows) of the entire spectrum in the LMG model derived from the algebra of angular momentum.
  We utilize distinctly colored symbols to distinguish states with distinct quantum numbers $l$.
  In different columns we have chosen different values for the longitudinal and transverse fields $\l_x$ and $\l_z$.}
  \label{FigureCTRTQFIfromAAM}
\end{figure}

Similar to the calculations of sections~\ref{sectionconcurence}, \ref{sectiontangle}, \ref{sectionQFI}, we utilize the expectation values of $\langle J_\alpha \rangle$ and $\langle J_\alpha J_\beta \rangle$ with $\alpha,\beta=x,y,z$ to compute the concurrence between two arbitrary sites, the tangle and residual tangle between one site and the other sites, and the quantum Fisher information.
The permutation invariance is necessary for the calculations of the concurrence and residual tangle, and they are only calculated for the eigenstates in the $l=\frac{N}{2}$ sector.
To calculate the tangle and quantum Fisher information, only translation invariance is required, and then we can compute them for all the eigenstates.
For small $N$, such as $N\leq16$, our results align with the ones obtained through bootstrap.
Furthermore, we are able to obtain results for significantly larger $N$ values, as demonstrated in Figure~\ref{FigureCTRTQFIfromAAM}.
The results for the larger $N$ values exhibit a qualitatively similar trend to the results obtained for smaller $N$ values via bootstrap.

\section{Solving LMG model from block diagonalization} \label{sectionLMGbd}

The LMG model with the Hamiltonian can also be solved through exact block diagonalization, as described in \cite{Sandvik:2010lkj}.
The Hamiltonian (\ref{HLMG}) exhibits translational invariance, and the momentum is a conserved quantity.
In the sector possessing a fixed momentum $k$, we are able to find the simultaneous eigenstates of the Hamiltonian and the angular momentum $J^2=J_x^2+J_y^2+J_y^2$.
By construction, every state displays translational invariance, although it may not necessarily be permutation invariant.
However, all the states with angular quantum number $l=N/2$ are permutation invariant.
For states with $l<N/2$, there exists a degeneracy among states with fixed $E$, $k$, and $l$, resulting in ambiguous wavefunctions for these states.

\begin{figure}[p]
  \centering
  \includegraphics[width=0.98\textwidth]{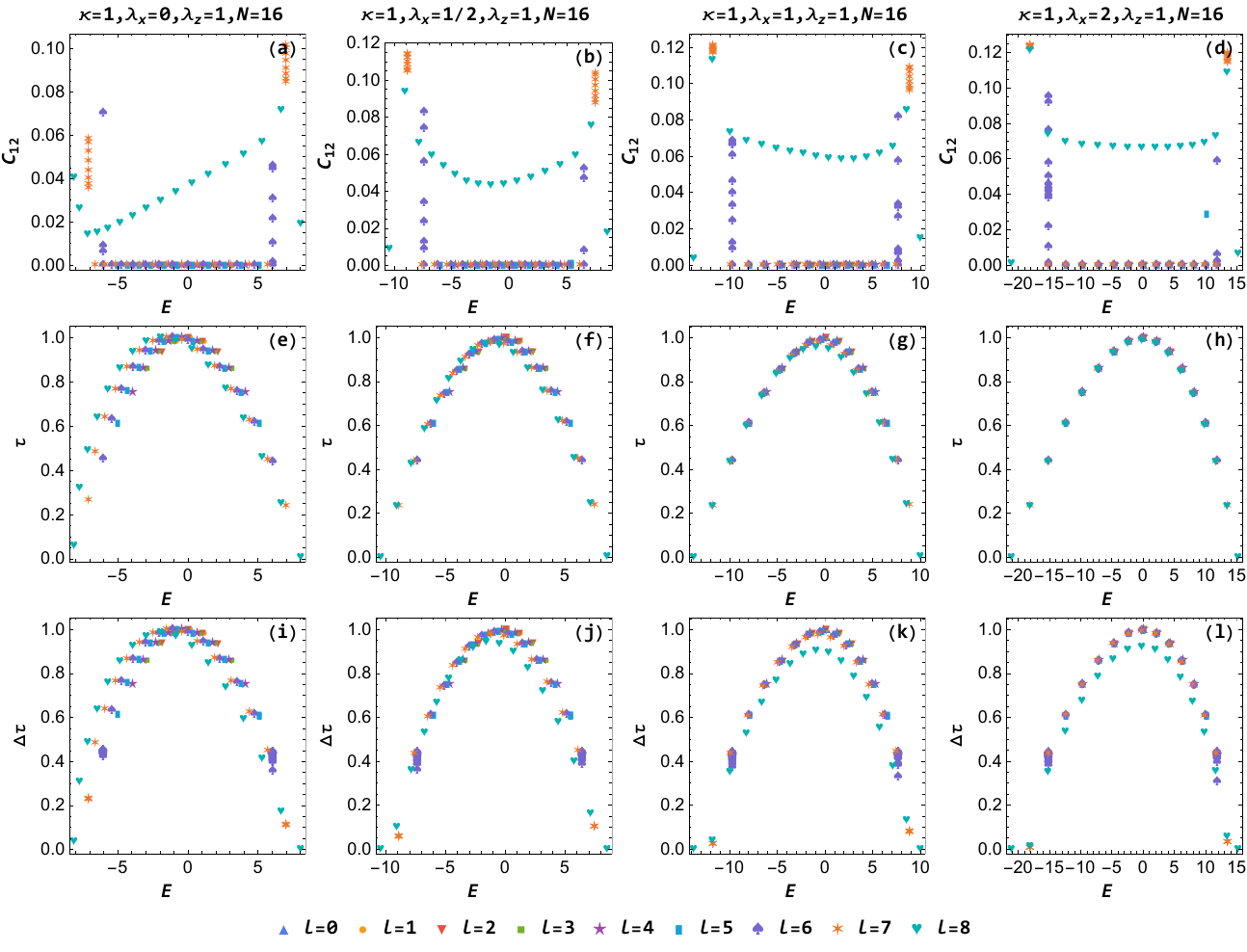}
  \caption{
  The concurrences between two nearby sites $C_{12}$ (first row), the tangle (second row) and the residual tangle (last row) between one site with other sites in the entire spectrum of the LMG model derived from block diagonalization.
  We utilize distinctly colored symbols to distinguish states with distinct quantum numbers $l$.
  In different columns we have chosen different values for the longitudinal and transverse fields $\l_x$ and $\l_z$.}
  \label{FigureCTRTfromBD}
\end{figure}

For each state, we compute the concurrences $C_{j_1j_2}$ with $(j_1,j_2) = (1,2), (1,1+\lfloor{N}/{4}\rfloor), (1,1+\lfloor{N}/{2}\rfloor)$ as well as the tangle and residual tangle between one site and other sites.
In Figure~\ref{FigureCTRTfromBD}, we present examples of the results for $N=16$.
For states with $l=N/2$, the outcomes align with those attained from bootstrap and angular momentum algebra.
However, for other states, disparities may arise from the arbitrary definition of the states, leading to differences from the results obtained through bootstrap and angular momentum algebra.

\section{Nonexistence of permutation invariant states in sector $l<N/2$} \label{sectionPIS}

In this appendix, we show that there are no permutation invariant states in the sector $l<N/2$ for $N>2$.

We first review briefly the relevant basics of the permutation group $S_N$. One could see more of permutation group in, for example, \cite{Dresselhaus:2008uzo}.
The permutation group $S_N$ has $N!$ elements, each of the which can be uniquely written as a product of several cyclic permutations, and elements with the same cyclic structure belong to the same conjugacy class.
An integer $N$ has $p_N$ different partitions, and each partition corresponds to one conjugacy class of $S_N$.
We denote a general partition of the integer $N$ as
\be \label{Ki}
K_i = m_1^{d_1} m_2^{d_2} \cdots m_{r_i}^{d_{r_i}}, ~ i=1,2,\cdots,p_N,
\ee
which indicates that the integer $m_s$ has $d_s$ duplicates for $s=1,2,\cdots,r_i$.
There are $1\leq m_1 < m_2 < \cdots < m_{r_i} \leq N$ and $\sum_{s=1}^{r_i} d_s m_s = N$.
The number of elements in the conjugacy class $K_i$ is
\be
n_i = \f{N!}{\prod_{s=1}^{r_i} ( m_s^{d_s} d_s! ) }.
\ee
The group $S_N$ with $N\geq2$ has only two one-dimensional representations: the symmetric representation $S(\cP)=1$ for $\forall\cP\in S_N$, and the antisymmetric representation $A(\cP)=1$ for an even permutation $\cP$ and $A(\cP)=-1$ if $\cP$ is an odd permutation.
For the conjugacy class $K_i$, the character of the one-dimensional symmetric and anti-symmetric representations are respectively
\be
\chi_S(K_i) = 1, ~ \chi_A(K_i) = \prod_{s=1}^{r_i} (-1)^{d_s(m_s-1)}.
\ee
The inner product of the characters of two representations are defined as
\be
\lag \chi | \chi' \rag \equiv \f{1}{N!} \sum_{i=1}^{p_N} n_i [\chi(K_i)]^* \chi'(K_i).
\ee
For $N\geq2$, one can easily check the orthonormality of the one-dimensional symmetric and anti-symmetric representations
\bea
&& \lag \chi_S | \chi_S \rag = \lag \chi_A | \chi_A \rag = \sum_{i=1}^{p_N} \prod_{s=1}^{r_i} \f{1}{m_s^{d_s}d_s!} = 1, \nn\\
&& \lag \chi_S | \chi_A \rag = \sum_{i=1}^{p_N} \prod_{s=1}^{r_i} \f{(-1)^{d_s(m_s-1)}}{m_s^{d_s}d_s!} = 0.
\eea

The $N$ spins form a Hilbert space $\cH$ of dimension $2^N$, which gives a representation of the permutation group $S_N$.
A permutation invariant state in the Hilbert space forms a one-dimensional representation of the permutation group $S_N$, which is either the symmetric representation $S$ or the antisymmetric representation $A$.
For the conjugacy class $K_i$ (\ref{Ki}), the character is
\be
\chi_\cH(K_i) = \prod_{s=1}^{r_i} 2^{d_s}.
\ee
We get the repetitions of the one-dimensional symmetric representation $S$ and antisymmetric representation $A$ in the representation $\cH$
\bea
&& \lag \chi_S | \chi_\cH \rag = \sum_{i=1}^{p_N} \prod_{s=1}^{r_i} \f{2^{d_s}}{m_s^{d_s}d_s!} = N+1, \nn\\
&& \lag \chi_A | \chi_\cH \rag = \sum_{i=1}^{p_N} \prod_{s=1}^{r_i} \f{(-1)^{d_s(m_s-1)}2^{d_s}}{m_s^{d_s}d_s!} =
    \lt\{ \ba{cc} 1, & N=2 \\ 0, & N \geq 3 \ea \rt.\!\!\!.
\eea
In appendix~\ref{sectionLMGangmom}, we see that there are already $N+1$ permutation invariant states in sector $l=\f{N}{2}$ in the Hilbert space.
Hence, we conclude that there are no permutation invariant states in the sectors with $l<\f{N}{2}$ for $N>2$.

\section{Computational complexity of the bootstrap method} \label{sectionCC}

In this appendix, we analyze the computational complexity of the bootstrap method.

For the LMG model (\ref{HLMG}) with fixed number of qubits $N$, we have chosen the space $\cS$ generally of $\pp N^3$ operators (\ref{cOabc}).
The number of linear constraint equations (\ref{newcon2}), (\ref{newcon3}), (\ref{newcon4}) and (\ref{newcon5}) are also proportional to $N^3$ and are easy to solve.
The major computational complexity resides at the evaluation of the tensor $g_{\a\b\g}$ (\ref{gabg}) from the matrix $B_{\a\b}$ and the tensor $C_{\a\b\g}$ (\ref{BabCabg}).
Each of the operator $\cO_\a$ is a $2^N \times 2^N$ matrix. To calculate $B_{\a\b}$ the number of traces we need to evaluate is proportional to $N^6$, and to calculate $C_{\a\b\g}$ the number of traces is proportional to $N^9$. This a quite heavy task for a large integer $N$.

In practice, we do not evaluate $g_{\a\b\g}$ from $2^N \times 2^N$ matrices.
We take the case that $N$ is an even integer as example.
Similar to the decomposition of the Hamiltonian (\ref{Hsim}), for each $\cO_\a$ we have the decomposition
\be
\cO_\a \sim \bigoplus_{l=0}^{N/2} [ \cO^{(l)}_\a ]_{d_{l}},
\ee
with the same duplicate number $d_l$ (\ref{dl}). For each $\cO_\a$ we define the corresponding operator
\be
\td \cO_\a = \bigoplus_{l=l_{\rm min}}^{N/2} \cO^{(l)}_\a,
\ee
which is just a $\f{(N+2)^2}{4} \times \f{(N+2)^2}{4}$ matrix.
We have
\be
\td \cO_\a \td \cO_\b = \sum_\g g_{\a\b\g} \td \cO_\g.
\ee
where the coefficients $g_{\a\b\g}$ are the same as the ones in (\ref{tensorgabg}).
Then we calculate $g_{\a\b\g}$ from
\be
g_{\a\b\g} = \sum_{\d} \td C_{\a\b\d} \td B^{-1}_{\d\g},
\ee
with the matrix $\td B_{\a\b}$ and tensor $\td C_{\a\b\g}$ being defined as
\bea
&& \td B_{\a\b} \equiv \tr(\td\cO_\a\td\cO_\b), \nn\\
&& \td C_{\a\b\g} \equiv \tr(\td\cO_\a\td\cO_\b\td\cO_\g),
\eea
and $\td B^{-1}$ being the inverse matrix of $\td B$.
In this way, we calculate the exact value of tensor $g_{\a\b\g}$ up to $N=16$.
Larger $N$ is possible, and it would take more time and storage space for the computer.

\providecommand{\href}[2]{#2}\begingroup\raggedright\endgroup

%\bibliographystyle{D:/00.bibx/JHEPx}
%\bibliography{D:/00.bibx/2023,D:/00.bibx/2022,D:/00.bibx/2021,D:/00.bibx/2020,D:/00.bibx/2019,D:/00.bibx/2018,D:/00.bibx/1960,D:/00.bibx/1970,D:/00.bibx/1980,D:/00.bibx/1990,D:/00.bibx/1995,D:/00.bibx/1996,D:/00.bibx/1997,D:/00.bibx/1998,D:/00.bibx/1999,D:/00.bibx/2000,D:/00.bibx/2001,D:/00.bibx/2002,D:/00.bibx/2003,D:/00.bibx/2004,D:/00.bibx/2005,D:/00.bibx/2006,D:/00.bibx/2007,D:/00.bibx/2008,D:/00.bibx/2009,D:/00.bibx/2010,D:/00.bibx/2011,D:/00.bibx/2012,D:/00.bibx/2013,D:/00.bibx/2014,D:/00.bibx/2015,D:/00.bibx/2016,D:/00.bibx/2017,D:/00.bibx/book,D:/00.bibx/work,D:/00.bibx/thesis}

%\newpage

\end{document}